\documentclass[submission,copyright,creativecommons]{eptcs}
\providecommand{\event}{ACL2 2022} % Name of the event you are submitting to
\usepackage{underscore}           % Only needed if you use pdflatex.

\usepackage{graphicx}
\usepackage{xcolor} 
\usepackage{amsmath}
\usepackage{amsthm}
\usepackage{amsfonts}
\usepackage{amssymb}
\DeclareMathOperator{\ord}{ord}
\newtheorem{thm}{Theorem}

\usepackage{listings}
\definecolor{commentsColor}{rgb}{0.497495, 0.497587, 0.497464}
\definecolor{keywordsColor}{rgb}{0.000000, 0.000000, 0.635294} %{1,0,0}
\definecolor{stringColor}{rgb}{0.558215, 0.000000, 0.135316}
\title{All Prime Numbers Have Primitive Roots}
\author{Ruben Gamboa
\institute{University of Wyoming \\ Laramie, Wyoming}
\email{ruben@uwyo.edu}
\and
Woodrow Gamboa
\institute{Stanford University \\ Stanford, California}
\email{woodrowg@stanford.edu}
}
\def\titlerunning{Primes Have Primitive Roots}
\def\authorrunning{R. \& W. Gamboa}
\begin{document}
\maketitle

\begin{abstract}
If $p$ is a prime, then the numbers $1, 2, \dots, p-1$ form a group under multiplication modulo $p$. A number $g$ that generates this group is called a primitive root of $p$; i.e., $g$ is such that every number between $1$ and $p-1$ can be written as a power of $g$ modulo $p$. Building on prior work in the ACL2 community, this paper describes a constructive proof that every prime number has a primitive root.
\end{abstract}

\section{Introduction}
\label{sec:intro}

This paper describes a proof in ACL2 of the fact that all prime numbers
have primitive roots. A \emph{primitive root} of a prime number $p$ is a
number $g$ such that all the numbers $1, 2, \dots, p-1$ can be written
as $g^n \bmod p$ for some value of $n$. For example, if $p=5$, then $g=2$
is a primitive root of $p$ since $1=2^4 \bmod 5$, $2=2^1 \bmod 5$, 
$3=2^3 \bmod 5$, and $4=2^2 \bmod 5$. However, for $p=7$, the number $2$
is not a primitive root of 7, because $2^n \bmod 7$ is always one of
$2$, $4$, or $1$. In particular, $2$ does not generate $3$ mod $7$. So not
all numbers in $1, 2, \dots, p-1$ are powers of $2$. The
reader can easily verify that $3$ is a primitive root of $7$, so the
theorem holds in this case.

More formally, if $p$ is a prime it is well known that the set of numbers
modulo $p$, written $\mathbb{Z}/p\mathbb{Z}$, forms a field. This occurs
because when $p$ is prime and for non-zero $a\in\mathbb{Z}/p\mathbb{Z}$, $a$ always has
a multiplicative inverse, i.e., a number $b\in\mathbb{Z}/p\mathbb{Z}$ such
that $ab\equiv1 \pmod p$. (Actually, inverses exist whenever $a$ and $p$ 
have no common factors, but this is guaranteed for all non-zero $a$ when $p$ is prime.)

The multiplicative group of this field, denoted by 
$\left(\mathbb{Z}/p\mathbb{Z}\right)^*$, contains the elements
$1, 2, \dots, p-1$ and $g$ is a primitive root of $p$ precisely when 
$g$ generates (in the sense of group theory) this group. So the fact that
prime numbers have primitive roots actually tells us something very interesting
about the structure of the group $\left(\mathbb{Z}/p\mathbb{Z}\right)^*$; 
it is a cyclic group, so it has the simplest possible structure.
Primitive roots also have applications to fast arithmetic modulo $p$, similar
to the way logarithms can be used to turn multiplication to addition over
the reals~\cite{silverman2006friendly}.

The ACL2 formalization of this result follows the hand proof presented
in~\cite{youtube:primitive-roots}.
The proof itself builds on two significant forays into
number theory in ACL2. First is Russinoff's proof of quadratic reciprocity,
which also defined the foundational notions of \texttt{divides},
\texttt{primep}, and useful lemmas such as that prime fields are 
integral domains (if $ab=0$ then either $a=0$ or $b=0$), and an important 
lemma due to Euclid (if $p$ divides $ab$, then either $p$ divides $a$ or $p$
divides $b$)~\cite{Russ:quadratic-reciprocity,Russ:quadratic-reciprocity-books}. We also built on top of 
Kestrel's formalization of prime fields, which includes definitions for the 
various field operations and their various arithmetic 
properties~\cite{kestrel:pfield}. As this brief list of prior results
indicates, many basic facts from number theory have already been formalized
in ACL2, but unfortunately the results are scattered in several places in
the community books. This is unfortunate, because number theory is a very
practical branch of mathematics, e.g., with applications to cryptography.
One thing we learned from this project is that it is time to collect these 
various results under a common branch of the community books, so that future 
projects can more easily build on top of the foundations that have already
been implemented.

The rest of this paper is organized as follows. In Sect.~\ref{sec:background},
we present some of the basic mathematical definitions in ACL2 of standard
concepts from number (and group) theory, like the order of a group element.
These are actually useful formalizations that could be used in other projects,
not just as part of this effort. Then in Sect.~\ref{sec:special-poly} we
discuss polynomial congruences, and in particular we prove that a special
family of polynomials have the greatest possible number of distinct roots.
This seemingly unrelated fact turns out to be a key technical lemma that is
used in Sect.~\ref{sec:order-construction} to construct elements
that have a desired order. The proof of the main theorem follows from these
constructions, and it is shown in Sect.~\ref{sec:primitive-root}. We conclude
the paper in Sect.~\ref{sec:conclusion} and give some ideas for future work.

\section{Mathematical Background}
\label{sec:background}

In this section, we discuss some mathematical foundations that are needed to
prove that all prime numbers have primitive roots. The definitions and proofs
in this section are very general and not solely for the purpose of our desired
theorem. In other words, these should be part of a global library of ACL2 books
formalizing number theory.

The first important concept is that of the order of an element of a group.
If $a\in\left(\mathbb{Z}/p\mathbb{Z}\right)^*$, the \emph{order} of $a$, denoted as
$\ord(a)$, is the least positive integer $k$ such that $a^{k}\equiv 1 \pmod p$.

The notion of order does not appear to be well-defined, since it seems possible that
$a^{k}\not\equiv 1 \pmod p$ for all positive integers $k$. But when $p$ is prime, an
important theorem of Fermat's says that this cannot be the case.
\begin{thm}[Fermat's Little Theorem]
	If $p$ is a prime number, and $a\in\left(\mathbb{Z}/p\mathbb{Z}\right)^*$, 
	then $a^{p-1}\equiv 1 \pmod p$.
\end{thm}
This theorem, formalized in ACL2 as part of~\cite{Russ:quadratic-reciprocity,Russ:quadratic-reciprocity-books},
immediately shows that $\ord(a) \le p-1$. We used this to define \texttt{order} in
ACL2. First, the function \texttt{(all-powers a p)} generates the list
$[a^1 \bmod p, a^2 \bmod p, \dots, a^k \bmod p]$ such that $k\le p-1$ and if $1\le i<k$, then
$a^i \bmod p \ne 1$. Clearly, the length of \texttt{(all-powers a p)} is between $1$ and $p-1$, 
inclusive, and when the length is less than $p-1$ the last element must be equal to 1.
Using Fermat's Little Theorem, it is easy to show that even when the length is exactly
equal to $p-1$, the last element is equal to 1. Then \texttt{(order a p)} is defined as
\texttt{(len (all-powers a p))}, and it follows that $a^{\ord(a)} \equiv 1 \pmod p$.

Another important theorem about order is that if $n$ is a positive integer such that 
$a^n \pmod p = 1$ and there does not exist a smaller positive integer $m$ such that
$a^m \pmod p = 1$, then in fact $\ord(a)=n$. We capture this theorem in ACL2 as
\begin{lstlisting}
(defthmd smallest-pow-eq-1-is-order
  (implies (and (fep a p)
                (not (equal 0 a))
                (primep p)
                (posp n)
                (equal (pow a n p) 1)
                (not (exists-smaller-power-eq-1 a p n)))
           (equal (order a p) n))
  :hints ...)
\end{lstlisting}
We include the ACL2 source of that theorem here, only to familiarize the reader with the functions
\texttt{fep} which recognizes elements of the field $\mathbb{Z}/p\mathbb{Z}$, \texttt{primep} which
recognizes primes, \texttt{pow} which performs exponentiation in the field, and its friends
\texttt{add}, \texttt{mul}, \texttt{inv}, etc., which perform the other arithmetic operations in 
the field---all of these were previously defined in the ACL2 Community Books.

An important fact about order is that for any element $a$, $\ord(a)$ divides $p-1$, which we will
write in the usual notation as $\ord(a) \mid p-1$. This follows because if $\ord(a)=n$, then the
list $L_n=[a^1 \bmod p, a^2 \bmod p, \dots, a^n \bmod p]$ ends in $1$, and $1$ does not appear
anywhere inside the list. But then $a^{n+k} \equiv a^n a^k \equiv a_k \pmod n$, so 
$L_{2n}=[a^1 \bmod p, a^2 \bmod p, \dots, a^{2n} \bmod p]$ is
simply two copies of $L_n$; i.e., $L_{2n} = app(L_n, L_n)$. That means that $L_{2n}$ ends in $1$,
and the only ones are $a^{n} \bmod p$ and $a^{2n} \bmod p$. This is easily extended to any multiple
of $n$, and since we know that $a^{p-1} \bmod p = 1$, it follows (almost) immediately that $p-1$
must be a multiple of $\ord(a)$, i.e., $\ord(a) \mid p-1$. This is actually a special case of
Lagrange's theorem for groups, but specialized for $\left(\mathbb{Z}/p\mathbb{Z}\right)^*$.

Another fact about \texttt{order} that is important to our proof is that the order of an inverse
is the same as the order of the element. I.e., $\ord(a^{-1}) = \ord(a)$. We proved this equality by
showing that both inequalities hold, and we used Lagrange's theorem to establish the inequalities.
The end result in ACL2 is as follows
\begin{lstlisting}
(defthmd order-inv
  (implies (and (fep a p)
                (not (equal 0 a))
                (primep p))
           (equal (order (inv a p) p) 
                  (order a p)))
  :hints ...)
\end{lstlisting}

We end this section by mentioning that the proof uses many facts about divides and the
greatest common divisor of two integers, formalized as \texttt{divides} and \texttt{g-c-d}
in~\cite{Russ:quadratic-reciprocity,Russ:quadratic-reciprocity-books}. And it also depends on many facts about the arithmetic
functions in $\mathbb{Z}/p\mathbb{Z}$, which were formalized in~\cite{kestrel:pfield}. While we 
needed to prove a handful of additional properties about many of these
these functions, the existing formalizations had already established most of the foundational
results, so this was mostly a matter of engineering the lemmas needed for our proof.

\section{A Special Polynomial Congruence}
\label{sec:special-poly}

In this section, we take an aside to consider polynomials modulo $p$. That is, we explore the roots
of polynomial congruences, such as
$$a_0 + a_1 x + \cdots + a_{n-1} x^{n-1} + a_n x^n \equiv 0 \pmod p.$$
The reason that polynomials pop up on a paper about prime numbers, is that polynomials can be used
as an alternative language to describe properties of congruences. For example, Fermat's Little
Theorem can be restated by saying that the polynomial congruence
$$-1 + x^{p-1} \equiv 0 \pmod p$$
has exactly $p-1$ distinct roots in $\left(\mathbb{Z}/p\mathbb{Z}\right)^*$.

Polynomials in ACL2 were formalized in~\cite{DBLP:conf/itp/GamboaC12} (among possibly many others),
but there are significant differences between polynomials and polynomial congruences. For 
example, the polynomial $x^2+2$ has no roots among the reals, but the similar polynomial congruence
$x^2+2$ does have a root in $\mathbb{Z}/11\mathbb{Z}$, because when $x=3$, 
$x^2+2=3^2+2=11\equiv 0 \pmod {11}$. So many of the properties of polynomials could not be
trivially transferred to polynomial congruences, and they had to be reproved from first principles.

One important lemma is that if $x$ is a root of the product of polynomials \texttt{poly1} and
\texttt{poly2}, then $x$ must be a root of at least one of those polynomials. This result depends
crucially on the fact that $\left(\mathbb{Z}/p\mathbb{Z}\right)^*$ is an integral domain when
$p$ is prime; i.e., if $ab\equiv0 \pmod p$ then either $a\equiv0 \pmod p$ or $b\equiv0 \pmod p$.
This lemma has a (mostly) immediate corollary, that the number of distinct roots of
the product of \texttt{poly1} and \texttt{poly2} is at most the number of distinct roots of
\texttt{poly1} plus the number of distinct roots of \texttt{poly2}. Note that all the roots must
be in $\left(\mathbb{Z}/p\mathbb{Z}\right)^*$, so the number of distinct roots of any polynomial
is at most $p-1$. This also means that it is possible to find a root methodically, by testing if $1$
is a root, or $2$ is a root, an so on. So if we know that a polynomial has a root, finding 
that root is guaranteed.

As a special case, consider a linear polynomial of the form $a_0 + a_1 x$, where $a_1 \not\equiv 0 \pmod p$.
Then $a$ is a root of this polynomial congruence if and only if $a = -a_0/a_1 \bmod p$, or in ACL2
\begin{lstlisting}
(defthm root-of-linear-pfield-polynomial
  (implies (and (primep p)
                (non-trivial-pfield-polynomial-p poly p)
                (equal (len poly) 2)
                (fep a p))
           (equal (pfield-polynomial-root-p poly a p)
                  (equal a (neg (div (first poly)
                                     (second poly)
                                     p)
                                p))))
  :hints ...)
\end{lstlisting} 
In particular, since the arithmetic operations return a single value, this also shows that a non-trivial
linear polynomial has exactly one root, where by ``non-trivial'' we mean that $a_1 \not\equiv 0 \pmod p$.

Now consider a general polynomial $P(x) = a_0 + a_1 x + \cdots + a_{n-1} x^{n-1} + a_n x^n$ with
$a_n \not\equiv 0 \pmod p$. Suppose that $a$ is a root of this polynomial. Then using the long-division
algorithm for polynomials, we can factor $P(x)$ into $P(x) = (x-a) Q(x)$ where 
$Q(x) = b_0 + b_1 x + \cdots + b_{n-2} x^{n-2} + b_{n-1} x^{n-1}$, for some suitable choice
of $b_i$.
\begin{lstlisting}
(defthm eval-poly-with-root
  (implies (and (integer-polynomial-p poly)
                (primep p)
                (integerp a)
                (fep x p)
                (pfield-polynomial-root-p poly a p))
           (equal (eval-pfield-polynomial poly x p)
                  (mul (eval-pfield-polynomial
                        `(,(- a) 1)
                        x p)
                       (eval-pfield-polynomial
                        (cdr (divide-polynomial-with-remainder-by-x+a
                              poly
                              (- a)))
                        x p)
                       p)))
  :hints ...)
\end{lstlisting}
This also shows that if $b$ is a root of $P(x)$, then either $b=a$ or $b$ is a root of $Q(x)$. In other
words, the number of distinct roots of $P(x)$ is at most 1 more than the number of distinct roots of
$Q(x)$.
If $n=1$, we've already seen that $P(x)$ has exactly one root in 
$\left(\mathbb{Z}/p\mathbb{Z}\right)^*$. So by induction, the number of roots of $P(x)$ is at most $n$.
\begin{lstlisting}
(defthm num-roots-of-poly-upper-bound
  (implies (and (primep p)
                (non-trivial-pfield-polynomial-p poly p)
                (<= 2 (len poly)))
           (<= (pfield-polynomial-num-roots poly p)
               (len (cdr poly))))
  :hints ...)	
\end{lstlisting}
This is, of course, a familiar and expected result for polynomials over the reals, but it is somewhat
surprising over $\mathbb{Z}/p\mathbb{Z}$, since it is possible that $P(a)\ne0$ but that $P(a)\equiv 0 \pmod p$.
I.e., it is possible that $a$ is a root of the congruence, but not of the polynomial over the reals. Nevertheless,
the total number of roots for the congruence is still bounded by $n$.

Now we introduce a special class of polynomials, which we call Fermat polynomials. The function
\texttt{(fermat-poly n)} constructs the polynomial $-1 + x^n$. Now suppose that $n=p-1$.
From Fermat's Little Theorem, it follows that this polynomial has exactly $n=p-1$ roots.
\begin{lstlisting}
(defthm num-roots-of-fermat-poly
  (implies (primep p)
           (equal (pfield-polynomial-num-roots (fermat-poly (1- p)) p)
                  (1- p)))
  :hints ...)	
\end{lstlisting}

Now, suppose that $n$ is a composite that can be written as $n=cd$, and again consider the polynomial $-1 + x^n$.
We observe that this polynomial can always be factored as
\begin{equation}
-1 + x^n = -1 + x^{cd} = (-1 + x^d)(1+x^d+x^{2d}+\dots+x^{(c-1)d}).
\label{eqn-1}	
\end{equation}
This result is easily proved on paper by expanding the right-hand side and matching up exponents. In ACL2, this is
a more technical proof that is really more about list manipulation. In particular, notice that the second polynomial
on the right-hand side consists of $c-1$ copies of the polynomial $x^d$ and that multiplying is by $x^d$ (and taking
into account the leading 1) results in $c$ copies of $x^d$ with a leading 0 consed in front. Summing the negated
polynomial then cancels all but the last copy of $x^d$, so the result is $-1+x^{cd}$. We found it convenient that
reasoning about exponents was reduced to reasoning about \texttt{cons} and \texttt{append} $k$ times, at which ACL2
excels.

Looking at Eqn.~\ref{eqn-1}, we see that the left-hand side has exactly $n$ roots when $n=p-1$ since $p$ is prime.
But the right-hand side has at most $d+(c-1)d$ distinct roots. Since $d+(c-1)d=d+cd-d=cd=n$, we conclude that both
polynomials in the product of the right-hand side must have the maximum number of distinct roots. In particular,
the polynomial $-1+x^d$ must have exactly $d$ distinct roots. Recall that the only thing special about $d$ is that
is divides $n$, so we have proved the following important technical lemma:
\begin{lstlisting}
(defthm num-roots-fermat-poly-divisor-implicit
  (implies (and (posp d)
                (primep p)
                (divides d (1- p)))
           (equal (pfield-polynomial-num-roots (fermat-poly d) p) d))
  :hints ...)
\end{lstlisting}
We will use this lemma in the next section.

\section{Constructing Elements of Given Order in $\left(\mathbb{Z}/p\mathbb{Z}\right)^*$}
\label{sec:order-construction}

In this section, we show how we can construct an element that has a desired order in 
$\left(\mathbb{Z}/p\mathbb{Z}\right)^*$, possibly by using other elements with known smaller order.

For starters, suppose that $a$ has order $m$ and $b$ has order $n$. What is the order of $ab$? In general, there's
not much we can say; e.g., if $b=a^{-1}$ then $ab=1$ so its order is $1$. But when $m$ and $n$ are relatively prime,
that is $\gcd(m,n)=1$, it turns out that the order of $ab$ is equal to $mn$.

To see this, observe that $(ab)^{mn}\equiv1\pmod p$. This follows because
\begin{align*}
(ab)^{mn} &\equiv \left((ab)^m\right)^n \\
          &\equiv \left(a^m b^m\right)^n \\
          &\equiv \left(1 b^m\right)^n \\
          &\equiv \left(b^m\right)^n \\
          &\equiv b^{mn} \\
          &\equiv \left(b^n\right)^m \\
          &\equiv 1^m \\
          &\equiv 1 \pmod p .
\end{align*}
As seen in Sect.~\ref{sec:background}, this implies that $\ord(ab) \mid mn$, which means $\ord(ab) \le mn$.
That is, $\ord(ab) \le \ord(a)\ord(b)$.

Now, suppose that $k$ is such that $(ab)^k\equiv1 \pmod p$. It follows that 
$a^k \equiv b^{-k} \equiv (b^{-1})^k \pmod p$. Raising both sides to the power $n$, we have that
$a^{nk} \equiv (b^{-1})^{nk}$. Since $\ord(b^{-1})=\ord(b)=n$, $(b^{-1})^{nk} \equiv 1 \pmod p$, so
$a^{nk} \equiv 1 \pmod p$ as well. This means that $\ord(a^k) \mid m$ and $\ord(a^k) \mid n$, and since
$\gcd(m,n)=1$ this means that the only possible value of $\ord(a^k)$ is 1.

All that is to show that $a^k \equiv b^k \equiv 1 \pmod p$. but that means that $m \mid k$ and $n \mid k$.
Again, since $\gcd(m,n)=1$ this means that $mn \mid k$. The only constraint on $k$ is that $(ab)^k\equiv1 \pmod p$,
so $\ord(ab)$ is such a $k$. This means that $\ord(a)\ord(b) \mid \ord(ab)$, so $\ord(a)\ord(b) \le \ord(ab)$.
Combined with the earlier inequality this shows that $\ord(ab)=\ord(a)\ord(b)$. In particular, given $a$ and $b$
with orders $m$ and $n$ that are relatively prime, this shows that we can construct an element with order $mn$:
\begin{lstlisting}
(defthm construct-product-order
  (implies (and (primep p)
                (fep a p) 
                (not (equal 0 a))
                (fep b p) 
                (not (equal 0 b))                
                (relatively-primep (order a p) (order b p)))
           (equal (order (mul a b p) p)
                  (* (order a p) (order b p))))
  :hints ...)
\end{lstlisting}
 
We now show how to construct an element that has a different special order. In particular, we wish to show that if
$p$ and $q$ are primes and $q^k \mid n=p-1$, then there is some element 
$g_{q^k}$ of $\left(\mathbb{Z}/p\mathbb{Z}\right)^*$ with order $q^k$.

We define the function \texttt{(number-of-powers x q)} which returns the largest power $k$ such that $q^k \mid x$.
For instance, \texttt{(number-of-powers 40 2)} is 3, since $40=2^3\cdot5$. Now suppose that $x$ divides a prime
power $q^n$. Then in fact, $x$ must be one of $1$, $q$, $q^2$, \dots, $q^n$. In particular, $x=q^k$ where $k$ is
the number of powers of $q$ in $x$ (and note that $k\le n$):
\begin{lstlisting}
(defthm factors-of-prime-powers
  (implies (and (primep q)
                (posp x)
                (natp n)
                (divides x (expt q n)))
           (equal x (expt q (number-of-powers x q))))
  :hints ...)
\end{lstlisting}

So suppose that $x$ is such that $x^{q^n}\equiv 1 \pmod p$, assuming for now that such an $x$ exists. Then clearly
$\ord(x) \mid q^n$, which means that $\ord(q^n)$ must be one of $1$, $q$, $q^2$, \dots, $q^n$. Now suppose also
that the order of $x$ is $q^i$ where $i<n$. Then $x^{q^i}\equiv 1 \pmod p$, so $x^{q^j}\equiv 1 \pmod p$ for any $j>i$. This follows because
\begin{align*}
x^{q^j} &\equiv x^{q^{i+j-i}} \\
        &\equiv x^{q^i q^{j-i}} \\
        &\equiv \left(x^{q^i}\right)^{q^{j-i}} \\
        &\equiv 1^{q^{j-i}} \\
        &\equiv 1 \pmod p
\end{align*}
In particular, if the order of $x$ is $q^i$ where $i<n$, it must be the case that $x^{q^{n-1}}\equiv 1 \pmod p$.
\begin{lstlisting}
(defthm order-is-prime-power-lemma
  (implies (and (primep p)
                (primep q)
                (<= q p)
                (fep a p)
                (not (= 0 a))
                (natp n)
                (= (pow a (expt q n) p) 1)
                (not (= (pow a (expt q (- n 1)) p) 1)))
           (equal (order a p) (expt q n)))
  :hints ...)
\end{lstlisting}
Now we address the question of whether such an $x$ exists. I.e., is there an $x$ such that both of these
equations hold:
\begin{align}
x^{q^n}& \equiv 1 \pmod p \label{eqn-2} \\
x^{q^{n-1}}& \not\equiv 1 \pmod p \label{eqn-3}
\end{align}
Note that for such an $x$, $\ord(x)$ is necessarily equal to $q^n$.

This is where the theorems about polynomials proved in Sect.~\ref{sec:special-poly} come into play. Eqn.~\ref{eqn-2}
holds precisely when $x$ is a root of the polynomial congruence $x^{q^n}-1\equiv 0 \pmod p$, and the theorem from
Sect.~\ref{sec:special-poly} guarantees that there are precisely $q^n$ distinct roots of this polynomial congruence,
as long as $q^n \mid p-1$. So there are $q^n$ values of $x$ that satisfy Eqn.~\ref{eqn-2}. Similarly, there are
$q^{n-1}$ values of $x$ that satisfy Eqn.~\ref{eqn-3}, again under the assumption that $q^{n-1} \mid p-1$, which is guaranteed when $q^n \mid p-1$. Since
$q^n > q^{n-1}$, there must be at least one $x$ that satisfies Eqn.~\ref{eqn-2} but not Eqn.~\ref{eqn-3}. It follows,
then that $\ord(x) = q^n$ for this particular $x$. Moreover, as observed earlier, the roots of any non-trivial
polynomial congruence must be one of $1$, $2$, \dots, $p-1$, so it is possible to \emph{find} an appropriate value
of $x$ by searching.
\begin{lstlisting}
(defthm order-is-prime-power
  (implies (and (primep p)
                (primep q)
                (natp n)
                (divides (expt q n) (1- p)))
           (and (fep (witness-with-order-q^n q n p) p)
                (not (= 0 (witness-with-order-q^n q n p)))
                (equal (order (witness-with-order-q^n q n p) p)
                       (expt q n))))
  :hints ...)
\end{lstlisting}
Using the two theorems proved in this section, we will show in the next how to find an element with order $p-1$, i.e.,
a primitive root of $p$.

\section{A Primitive Root of $p$}
\label{sec:primitive-root}

Using the results proved in Sect.~\ref{sec:order-construction}, it is straightforward to
prove that all prime numbers have primitive roots. The typical pen-and-paper proof goes
like this. Suppose that $p$ is an odd prime. (If $p=2$, it is obvious that $1$ is a primitive
root.) Factor the number $p-1$ as a product of prime powers, as in
$$p-1 = {q_1}^{k_1} \cdot {q_2}^{k_2} \cdot \dots  \cdot {q_m}^{k_m}.$$
Now, for each term ${q_i}^{k_i}$, there is an element $c_i$ of order  ${q_i}^{k_i}$.
Note that all the $q_i$ are primes distinct from one another, so the gcd of any ${q_i}^{k_i}$ 
and any product of other ${q_j}^{k_j}$ must be $1$. So the $c_i$ are numbers of order
${q_i}^{k_i}$ which are relatively prime. So $c=c_1\cdot c_2 \cdot \dots \cdot c_m$ must have
order ${q_1}^{k_1} \cdot {q_2}^{k_2} \cdot \dots  \cdot {q_m}^{k_m} = p-1$.  Thus $c$ is
a primitive root of $p$.

We could have followed this approach in ACL2, and in fact prime
factorization has been formalized in ACL2 and NQTHM numerous times, 
e.g., in~\cite{DBLP:conf/acl2/CowlesG06}. But this turned out not to
be very helpful for two reasons. First, the formalization 
in~\cite{DBLP:conf/acl2/CowlesG06} uses a different (albeit
equivalent) definition of ``prime.'' This is a common situation in
the ACL2 formalizations of number theory, and it is something that we
would like to see addressed. Second, the result about primitive roots
does not depend on the full Fundamental Theorem of Arithmetic; i.e.,
what we need is that the number $p-1$ can be decomposed into prime
powers, but we do not need that the decomposition is unique.
Naturally, the uniqueness property is the hardest part of the
proof. So simply proving a weak version of prime factorization would
be easy and effective for our purposes.

In fact, it's possible to decompose $p-1$ into powers of primes and
compute the primitive root $c$ at the same time. The first step is to
define the function \texttt{(primitive-root-aux k p)} that finds
an element of order \texttt{k}:
\begin{lstlisting}
(defun primitive-root-aux (k p)
  (if (or (zp k) (= 1 k))
      1
    (let* ((q (least-divisor 2 k))
           (n (number-of-powers k q))
           (k1 (/ k (expt q n))))
      (mul (witness-with-order-q^n q n p)
           (primitive-root-aux k1 p)
           p))))
\end{lstlisting}
The difficult part is helping ACL2 admit this function by proving
that it always terminates. The function \texttt{least-divisor},
defined in~\cite{Russ:quadratic-reciprocity,Russ:quadratic-reciprocity-books}, finds the smallest
divisor (starting at 2) of \texttt{k}. For $k\ge2$, it is shown
in~\cite{Russ:quadratic-reciprocity,Russ:quadratic-reciprocity-books} that this number is always a
prime less than or equal to $k$. Then the function 
\texttt{number-of-powers} finds the corresponding exponent in the
prime decomposition of \texttt{k}. Using those facts, we proved that 
$k/q^n$ is a natural number that is smaller than $k$, thus proving
the termination of the function.

By inspection, it is easy to see that this function does return the
primitive root $c$ described in the hand proof. We proved that in
ACL2 using an induction suggested by the function
\texttt{primitive-root-aux} to prove the following key theorem:
\begin{lstlisting}
(defthm primes-have-primitive-roots-aux
  (implies (and (primep p)
                (natp k)
                (divides k (1- p)))
           (equal (order (primitive-root-aux k p) p)
                  k))
  :hints ...)	
\end{lstlisting}
In order for this to work, we had to prove a number of technical
lemmas. For starters, since we're using induction as suggested by
\texttt{primitive-root-aux} we have to show that $k/q^n$ is
a natural that divides $p-1$ whenever $k$ is a natural that divides
$p-1$. And we also had to show that the two terms multiplied in
\texttt{primitive-root-aux} satisfy the conditions of the theorems
\texttt{construct-product-order} and 
\texttt{order-is-prime-power} that are used to create the element
of order $k$. The most interesting are
\begin{itemize}
\item  the functions return elements in the multiplicative group
$\left(\mathbb{Z}/p\mathbb{Z}\right)^*$, 
\item in particular the result of those operations is never $0$,
\item the number $k/q^n$ divides $p-1$ if $k$ divides
$p-1$, 
\item and the gcd of $q^n$ and $k/q^n$ is $1$.	
\end{itemize}
Once that is done, the primitive root of $p$ can be defined and
shown to be a primitive root as follows:
\begin{lstlisting}
(defund primitive-root (p)
  (primitive-root-aux (1- p) p))

(defthm primes-have-primitive-roots
  (implies (primep p)
           (equal (order (primitive-root p) p)
                  (1- p)))
  :hints ...)	
\end{lstlisting}

\section{Conclusion}
\label{sec:conclusion}

In this paper, we presented a proof that all prime numbers have at least one primitive root. 
In fact, the number of primitive roots of $p$ can be shown to be $\phi(p-1)$ where $\phi$ is
Euler's totient function (the number of positive integers up to $n$ that are relatively prime
to $n$). Proving that would be a nice extension to this work that could happen in the future.

The proof relied on prior work on number theory, but our experience suggests that the prior
work is scattered across many directories in the community books. Moreover, many foundational
results needed to be proved to complement the existing foundations. This reflects the fact that
the development of number theory in ACL2 has been driven by specific results, so the foundations
developed in each project are tailored to support the needs of those specific projects. Given the
importance of number theory in areas such as cryptography, as well as the suitability of ACL2 to
reason effectively about this branch of mathematics, we think is would be a great time to consolidate
these formalizations in ACL2 under a common location in the community books.
Recent discussions in the ACL2 mailing list suggest that there is enough momentum to carry
out this project.

\nocite{*}
\bibliographystyle{eptcs}
\bibliography{primitive-roots}
\end{document}